\documentclass[prl,twocolumn,showpacs,superscriptaddress]{revtex4-1}
\usepackage{graphicx,amssymb,amsmath,color}
\usepackage{hyperref}

\begin{document}
\definecolor{darkgreen}{rgb}{0,0.5,0}

\title{The Kibble-Zurek mechanism at exceptional points}

\author{Bal\'azs D\'ora}
\email{dora@eik.bme.hu}
\affiliation{Department of Theoretical Physics and MTA-BME Lend\"ulet Topology and Correlation Research Group,
Budapest University of Technology and Economics, 1521 Budapest, Hungary}
\author{Markus Heyl}
\affiliation{Max-Planck-Institut f\"ur Physik komplexer Systeme, 01187 Dresden, Germany}
\author{Roderich Moessner}
\affiliation{Max-Planck-Institut f\"ur Physik komplexer Systeme, 01187 Dresden, Germany}

\date{\today}

\begin{abstract}
Exceptional points (EPs) are ubiquitous in non-hermitian systems, 
and represent the complex counterpart of  critical points.
By driving a system through a critical point at finite rate induces defects, 
described by the Kibble-Zurek mechanism, which finds applications in diverse fields of physics.
Here we generalize this to a ramp across an EP. 
We find that adiabatic time evolution brings the system
into an eigenstate of the final non-hermitian Hamiltonian and demonstrate that 
for a variety of drives through an  EP, the defect density
scales as $\tau^{-(d+z)\nu/(z\nu+1)}$ in terms of the usual critical exponents and $1/\tau$ the speed of the drive.
Defect production is suppressed compared to the conventional hermitian case as the defect state
can decay back to the ground state close to the EP. 
We provide a physical picture for the studied dynamics through a mapping onto a  Lindblad master equation with an additionally imposed continuous measurement.
\end{abstract}

\maketitle

The Kibble-Zurek mechanism of universal defect production is a paradigmatic phenomenon in nonequilibrium many-body physics~\cite{kibble,zurek,dziarmaga,polkovnikovrmp}.
While the adiabatic theorem ensures that a system can follow its ground state upon dynamically changing a Hamiltonian parameter sufficiently slowly as long as the spectrum remains gapped, this is no longer the case when crossing a continuous phase 
transition\cite{sachdev}.
As a consequence, excitations are generated and the number of defects $n$ exhibits universal behavior with a scaling that is determined solely by the universality class of the underlying phase transition:
\begin{equation}
	n \sim \tau^{-d\nu/(z\nu +1)} \, .
	\label{eq:KZscaling}
\end{equation}
Here, $1/\tau$ denotes the rate at which the parameter is dynamically varied, $d$ the spatial dimension, and $\nu$ and $z$ the correlation length and dynamical critical exponent, respectively.

The Kibble-Zurek mechanism in quantum many-body systems applies to unitary real-time evolution.
However, recent developments suggest rich features appearing for non-hermitian Hamiltonians describing intrinsically non-unitary dynamics~\cite{ptreview,rotter,bender98,berry2004}, as recently realized also in experiments~\cite{zeuner,gao2015}.
 While the eigenvalues of a non-hermitian 
Hamiltonian can still be interpreted in terms of energy bands, already the meaning of its eigenvectors cannot be treated conventionally as they are not orthogonal, and therefore 
possess finite overlap already in the absence of any additional perturbation.
Particularly important in this context are exceptional points~\cite{heiss} (EPs), where the complex spectrum becomes gapless. These can be regarded as 
the non-hermitian counterpart of conventional quantum critical points\cite{zhou18,sachdev}.
At EPs, two (or more) complex eigenvalues and \emph{eigenstates} coalesce, which then no longer form a complete basis.

In this work we study the defect production at EPs upon slowly changing a system parameter in the spirit of the Kibble-Zurek mechanism.
We identify a channel for defect suppression unique to EPs, absent from Hermitian dynamics.
Due to non-orthogonality of wavefunctions, only a small fraction of the excited state, which points perpendicular to the ground state, accounts for defect production.
Remarkably, however, while we find that the number of defects $n$ differs from the unitary Kibble-Zurek result in Eq.~(\ref{eq:KZscaling}), it still obeys a universal scaling form where $d$ in Eq. \eqref{eq:KZscaling} is replaced by a modified effective dimension $d_\mathrm{eff}=d+z$, involving also the dynamical critical exponent $z$ associated with the EP.

We study defect production for a set of different, but complementary, protocols of parameter ramps, which allows us to address different aspects of defect 
production in non-hermitian systems.
We provide a physical interpretation of our results in terms of an open quantum system described by a Lindblad master equation with an additionally imposed continuous measurement.

\section{Results}

\subsection{The model and observables}

We consider Hamiltonians of the form~\cite{dattoli,tonylee,carmichael,solano}
\begin{gather}
H=\sum_p H_p, \quad H_p=p\sigma_x+\Delta\sigma_y+i\Gamma\sigma_z
\label{hamilton}
\end{gather}
which can be decomposed into different momentum sectors labeled by $p$. For $i\Gamma \in  \mathbb{R}$ the problem is Hermitian.
For $\Gamma \in \mathbb{R}$ instead, the above Hamiltonian becomes non-hermitian with a spectrum given by $E_\pm(p)=\pm\sqrt{p^2+\Delta^2-\Gamma^2}$. 
When $\Delta>\Gamma$, $H_p$ has real eigenvalues for each $p$. For $\Gamma>\Delta$ on the other 
hand $H_p$, has, in general, complex eigenvalues. At sufficiently large $p>\sqrt{\Gamma^2-\Delta^2}$, however, the spectrum becomes real again.
A Hamiltonian is PT-symmetric if it commutes with the combined parity and time reversal operators.
Its spectrum is real if PT-symmetry is not spontaneously broken, i.e. the eigenstates also respect PT symmetry.
With broken PT-symmetry, the spectrum becomes complex. The analysis of higher order exceptional points~\cite{heiss}  is beyond the scope of the current investigation.
Non-Hermitian Hamiltonians of the kind in Eq. \eqref{hamilton} can be emulated by optical waveguides\cite{zeuner,ptreview,tonylee}, distributed-feedback structures\cite{longhi2010}, microcavities\cite{gao2015}
 or electric circuits\cite{stehmann}.
Eq. \eqref{hamilton} also accounts for the low energy dynamics of the quantum Ising chain in an imaginary transverse field\cite{complexising,deguchi} or an imaginary mass fermion (i.e. tachyon) system\cite{solano}. 
The last term in Eq. \eqref{hamilton} assumes balanced gain and loss without loss of generality: one can shift  the diagonal term in the Hamiltonian by any complex value without affecting the results (see Methods). 

\subsection{Hermitian dynamics} 

We investigate several natural scenarios for time-dependent $\Delta$ and $\Gamma$.
Let us start with the hermitian Kibble Zurek mechanism, which is contained in our model by 
choosing $\Delta=0$, $\Gamma=-i\Delta_0t/\tau$. This is the conventional Landau-Zener problem\cite{landau,zener} starting exactly from the critical point, thus it represents  only a half crossing. 
This yields (see Methods)
$\langle\sigma_y(\tau)\rangle=0$, while
\begin{gather}
\langle\sigma_z(\tau)\rangle+\frac{\Delta_0}{\pi}\ln\left({2W}/{\Delta_0}\right)\sim \tau^{-1/2},
\end{gather}
with $W$ the high energy cutoff.
Defect production is effective when the adiabatic condition is violated\cite{degrandi,polkovnikov}, namely when $d\ln|\Gamma|/d t\sim |\Gamma|$. 
This gives the transition time $\tau^{1/2}$, and the defect density scales inversely with this, in accord with Ref. \cite{bermudez2009}.
This also follows from the scaling behaviour of the matrix element
\begin{gather}
\sigma_z(p,\tau)=\sigma_z^\textmd{eq}(p)+f_\textmd{LZ}\left(\frac{p}{\Delta_0}(\tau\Delta_0)^{1/2}\right),
\label{hermitianscaling}
\end{gather}
which is typical for the hermitian Landau-Zener problem, $\sigma_z^\textmd{eq}(p)=-\Delta_0/\sqrt{p^2+\Delta_0^2}$. 
$f_\textmd{LZ}(x)$ is a universal scaling function in the near-adiabatic limit, which  decays exponentially with $x$.
For a general quantum critical point, the momentum resolved defect density is
\begin{gather}
n(p,\tau)=\tilde f_\textmd{LZ}\left(p^z\tau^{z\nu/(z\nu+1)}\right),
\end{gather}
and the defect density follows the Kibble-Zurek scaling\cite{kibble,zurek} as
\begin{gather}
n(\tau)=\int \frac{d^dp}{(2\pi)^d}~n(p,\tau)\sim \tau^{-d\nu/(z\nu+1)}
\end{gather}
with $d$, $z$ and $\nu$ being the spatial dimension, dynamical critical exponent and the exponent of the correlation length, respectively.

\subsection{Gapless quench} 

This is realized for $\Delta=\Gamma=\frac{\Delta_0}{2}t/\tau$. 
The eigenvalues of the Hamiltonian are always $\pm |p|$, irrespective of the value of $\Delta_0$, which makes this parameter ramp exactly solvable 
by plugging these parameters into the non-hermitian Schr\"odinger equation (see Methods). 
The dynamics is nevertheless non-trivial and
as $\Delta$ and $\Gamma$ evolve with time, defects are produced in spite of the fact that the instantaneous eigenvalues do not change.
Starting from the ground state at $t=0$, for $p<0$ the wavefunction only picks up a phase factor as $\exp(-ipt)[1,1]^T/\sqrt{2}$. 
On the other hand, the $p>0$ ground state at $t=0$ evolves to
\begin{gather}
\Psi_p(\tau)=\left[\begin{array}{c}
1-\frac{i\Delta_0}{2p}\\
-1-\frac{i\Delta_0}{2p}\end{array}\right]
\frac{\exp(ip\tau)}{\sqrt 2}+
\left[\begin{array}{c}
1 \\ 1
\end{array}\right]
\frac{i\Delta_0\sin(p\tau)}{2\sqrt 2 p^2\tau},
\end{gather}
where the second term is generated by the time dependence. 
For $\tau\rightarrow\infty$, this expression agrees with the right eigenfunction of the final Hamiltonian (up to normalization).
Not only does the instantaneous eigenvalue remain unchanged, i.e. $E_\pm(p)=\pm |p|$, but also the time evolution is only sensitive to the instantaneous eigenenergies, namely the wavefunction contains $\exp(\pm ipt)$ exponential factors only.
This is then used in Eq. \eqref{expvalue} to yield $\langle\sigma_y(\tau)\rangle+\Delta_0\ln(2W/\Delta_0)/2\pi\sim  \tau^{-2}$ and 
$\langle\sigma_z(\tau)\rangle\sim \tau^{-1}$.
In this case, the $\tau\rightarrow\infty$ solution coincides with the instantaneous expectation value after the time evolution and an adiabatic theorem seems to hold.
Since the instantaneous spectrum remains unchanged and gapless throughout, the above scaling cannot originate from the usual 
argumentation of Kibble-Zurek scaling.
This is analogous to quenching along a gapless line\cite{pellegrini} within the hermitian realm.

In order to appreciate the role of wavefunction normalization in Eq. \eqref{expvalue}, we have also 
evaluated it without the denominator: $\langle\sigma_z(\tau)\rangle$ approaches a constant, 
while $\langle\sigma_y(\tau)\rangle\sim\ln(\tau)$, without any well defined adiabatic limit for $\tau\rightarrow\infty$.

\subsection{PT-symmetric ramp} 

\begin{figure}
\includegraphics[width=6.5cm]{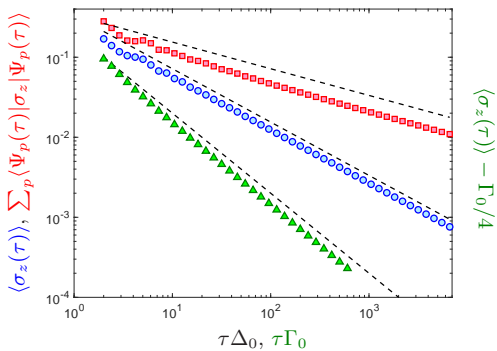}
\caption{The numerically determined defect density. We plot the defect density from
the normalized (blue circles) and unnormalized wavefunction (red squares) for the PT-symmetric ramp, as well as
for the full non-hermitian drives (green triangles), measured from its adiabatic value.
The black dashed lines denote the $\tau^{-1/3}$, $\tau^{-2/3}$ and $\tau^{-1}$ scaling.
The cutoff $|p|<W=10\Delta_0$ and $10\Gamma_0$, respectively, does not alter the dynamics, with other values yielding similar scaling.}
\label{figcase3}
\end{figure}

Now we consider a fully PT symmetric Kibble-Zurek problem, when the instantaneous spectrum is always real. We choose $\Delta=\Delta_0$, $\Gamma=\Delta_0t/\tau$ such that the time evolution ends exactly at an EP. 
The spin expectation values are $\langle\sigma_y(\tau)\rangle+\Delta_0\ln(W/\Delta_0)/\pi\sim \tau^{-2/3}$, and
\begin{gather}
\langle\sigma_z(\tau)\rangle\sim\tau^{-2/3},
\label{kz3}
\end{gather}
as shown in Fig. \ref{figcase3} from the numerics. The wavefunction for $t=\tau\rightarrow\infty$ agrees with the non-normalized right eigenfunction of the final non-hermitian
Hamiltonian, similarly to the gapless quench.

The gap in the instantaneous spectrum reads $\tilde\Delta=\Delta_0\sqrt{1-t^2/\tau^2}\approx\Delta_0\sqrt{2}\sqrt{(\tau-t)/\tau}$ for $t\sim\tau$. The distance from the critical 
point is $\hat t=\tau-t$, which is used to obtain the critical exponents $z\nu=1/2$ from the scaling
of the gap\cite{sachdev}, $\tilde\Delta\sim |\hat t|^{z\nu}$.
Then, the transition time\cite{degrandi,polkovnikov} separating a/diabatic dynamics is determined from $\tilde\Delta^2\sim d\tilde\Delta/ d\hat t$, which gives the 
transition time $\hat t_\textmd{tr}\sim \tau^{1/3}$,
in agreement with Kibble-Zurek scaling\cite{kibble,zurek} $t_\textmd{tr}\sim \tau^{z\nu/(z\nu+1)}$.
Note that similar critical exponents apply also to the Hermitian Rabi model\cite{plenio}. 

Since the spectrum $\pm |p|$ is linearly gapless at the critical point, this defines $z=1$, leaving us with $\nu=1/2$ for the exponent of the correlation
length. Therefore, the Kibble-Zurek scaling of the defect density in one dimension predicts $\sim \tau^{-d\nu/(z\nu+1)}=\tau^{-1/3}$ scaling. However, this exponent is different from Eq. \eqref{kz3}.
We demonstrate that the correct exponent is indeed $-2/3$ and present a generalized Kibble-Zurek scaling to account for that.

\begin{figure}
\includegraphics[width=6.5cm]{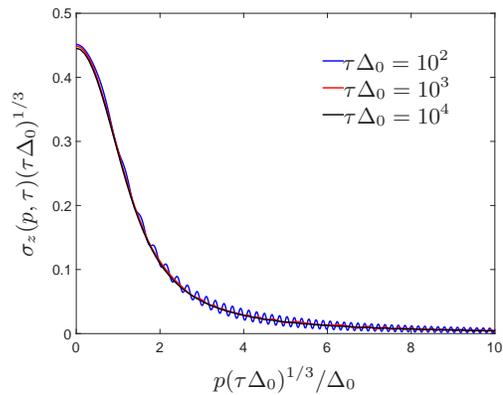}
\caption{Momentum resolved defect density for the PT-symmetric ramp.
The scaling and data collapse of the numerically determined momentum resolved defect density, $f_\textmd{PT}(x)$ in the near adiabatic limit is shown
for several values of $\tau$ for the PT-symmetric ramp.}
\label{scalingcase3}
\end{figure}

First of all, the numerical data indicates that the momentum resolved defect density, 
$\sigma_z(p,\tau)$,
 follows the scaling relation
\begin{gather}
\sigma_z(p,\tau)=\frac{1}{(\tau\Delta_0)^{1/3}}f_\textmd{PT}\left(\frac{p}{\Delta_0}(\tau\Delta_0)^{1/3} \right)
\label{scalingfunc3}
\end{gather}
with $f_\textmd{PT}(x)$ the universal scaling function shown in Fig. \ref{scalingcase3}. Upon 
integrating this with respect to $p$ by changing variable $x=p(\tau\Delta_0)^{1/3}/\Delta_0$, the $\tau^{-2/3}$ scaling of the defect density follows.
In Eq. \eqref{scalingfunc3}, 
the $\tau$ exponent 1/3 originates from the $z\nu/(z\nu+1)$ combination of critical exponents and the $p$ stems from the $z=1$ dynamical critical exponents. Therefore, this expression is 
generalized to an arbitrary critical point for the momentum resolved defect density as
\begin{gather}
n(p,\tau)=\frac{1}{\tau^{z\nu/(z\nu+1)}}\tilde f_\textmd{PT}\left(p^z\tau^{z\nu/(z\nu+1)} \right),
\label{scalingfunc3gen}
\end{gather}
which, after performing a $d$-dimensional momentum integral, gives $n\sim\tau^{-(d+z)\nu/(z\nu+1)}$.
Notice the $\tau$ dependent prefactor of the scaling functions in Eqs. \eqref{scalingfunc3} and 
\eqref{scalingfunc3gen} compared to the hermitian case Eq. \eqref{hermitianscaling}.

We next provide three complementary explanations for this modified scaling.
In an a/diabatic picture, excitations are created by populating the excited state similarly to hermitian dynamics, but 
only its component perpendicular to the ground state represents defect production.
 As we approach the EP with increasing time, we enter into  the diabatic regime at the transition time $\hat t_\textmd{tr}$, where 
adiabatic time evolution breaks down, the dynamics gets frozen and defect production kicks in.
The component of the excited state perpendicular to the ground state at this instance has an amplitude $\sin(\theta_p)$  as the ground and excited states are not orthogonal in general~\cite{berry}.
For Eq. \eqref{hamilton}, this is evaluated for small momentum states close to the EP, which are the most sensitive to diabatic time evolution,
as $\sin(\theta_{p\approx 0})=\sqrt{\Delta^2-\Gamma^2}/\Delta \sim 1/\hat t_\textmd{tr}$ at the adiabatic-diabatic transition: 
namely
the angle becomes proportional to the energy gap.
This results in a $\tau^{-z\nu/(z\nu+1)}$ suppression factor for the defect density. For the hermitian case, orthogonality ensures that $\sin(\theta)=1$.

In a more dynamical picture, defects are created directly in the state perpendicular to the ground state, which decays to the ground state with a 
rate $1/\hat t_\textmd{tr}$ reducing the Hermitian Kibble-Zurek scaling by the probability
to remain in the perpendicular state, $1/\hat t_\textmd{tr}$. At an EP, there is only a single eigenstate and any perpendicular component decays~\cite{berry94}.
Close to an EP, the state perpendicular to the ground state
initially decays towards the ground state, which is followed by revival and periodic oscillation with frequency $E_+(p)$.
 An unnormalized state perpendicular to the ground state evolves as 
$|\perp (t)\protect\rangle=2i \cot(\theta_p) \sin(E_+(p)t)|GS\protect\rangle+
\exp(iE_+(p)t)|\perp\protect\rangle$ due to a time independent non-hermitian Hamiltonian, where
$\theta_p$ is the angle between the ground state and the excited state, $\protect\langle GS|\perp\protect\rangle=0$ and $\theta_p$ vanishes at EP.
For $t\sim \pi/2E_+(p)$, it develops a large component parallel to the ground state with length $\cot(\theta_p)$. 
However, when 
the driving rate, $\partial_t E_+(p)/E_+(p)$ is larger than the revival frequency, the system does not have enough time for revival and only the initial decay is probed.
This gives the very same condition as the a/diabatic transition and the decay time is $\hat t_\textmd{tr}$.

Mathematically, the prefactor in Eq. \eqref{scalingfunc3gen} arises because the norm of the wavefunction also changes due to non-unitary time evolution. 
This follows from\cite{graefe2008}  
$d \langle\Psi_p(t)|\Psi_p(t)\rangle/dt=2\Gamma(t)\langle\Psi_p(t)|\sigma_z|\Psi_p(t)\rangle$. 
Due to the slow time evolution, states with large momentum $p$ have large energy, and are hardly affected by the time dependent term 
and the corresponding wavefunction norm hardly changes. The low energy and small momentum states are the most influenced by the non-hermitian and non-adiabatic 
time evolution. At short times ($t\ll\tau$), both the small matrix element and the $\Gamma(t)$ prefactor block the growth of the wavefunction norm, 
but at a distance $\hat t_\textmd{tr}$ from the critical 
point, adiabaticity breaks down. Afterwards, diabatic time evolution takes place, and the norm of the wavefunction gets enhanced by 
$\hat t_\textmd{tr}\sim \tau^{1/3}$, thus 
suppressing the defect density.

\subsection{Full non-hermitian drive} 

A full non-hermitian drive is realized for $\Delta=0$, $\Gamma=\Gamma_0t/\tau$, which represents the non-hermitian Kibble-Zurek problem and 
is equivalent to quenching the imaginary tachyon mass\cite{solano}. The instantaneous spectrum contains EPs located at $|p|=\Gamma$. 
By expanding around the EP, the spectrum 
scales as $E_\pm(p\gtrsim\Gamma)=\pm \sqrt{2\Gamma}\sqrt{p-\Gamma}$ in the PT symmetric 
regime, and as $E_\pm(p\lesssim\Gamma)=\pm i\sqrt{2\Gamma}\sqrt{\Gamma-p}$ in the broken PT symmetry sector.
Altogether the dynamical critical exponent is thus $z=1/2$, while $\nu=1$.
During the time evolution, all instantaneous eigenvalues are imaginary for $\Gamma(t)>|p|$  and real for $\Gamma(t)<|p|$, separated
by an EP. This critical point, which is located at $p=\Gamma_0t/\tau$,  moves in momentum space during the time evolution at a speed $\Gamma_0/\tau$, producing defects in the process. This results in
$\langle\sigma_y(\tau)\rangle=0$ and
\begin{gather}
\langle\sigma_z(\tau)\rangle-\Gamma_0/4\sim\tau^{-1} \, .
\end{gather}
This is depicted in Fig. \ref{figcase3} from the numerical solution of Eq. \eqref{timedepsch} (see Methods). 
Here, it is again crucial to properly normalize the wavefunction as in Eq. \eqref{expvalue}. 
Without the normalization, the spin expectation value changes exponentially in time due to the imaginary energy eigenvalues.

The numerically obtained adiabatic value for $\langle\sigma_z(\tau\rightarrow\infty)\rangle$ is also corroborated from diagonalizing 
the non-hermitian Hamiltonian analytically, and using its normalized right eigenfunction:
\begin{gather}
\langle\sigma_z\rangle_\textmd{eq}=\frac{1}{2\pi}\int\limits_{-\Gamma_0}^{\Gamma_0}dp\frac{\sqrt{\Gamma_0^2-p^2}}{\Gamma_0}=\frac{\Gamma_0}{4}.
\end{gather}

\begin{figure}[t]
\includegraphics[width=6.5cm]{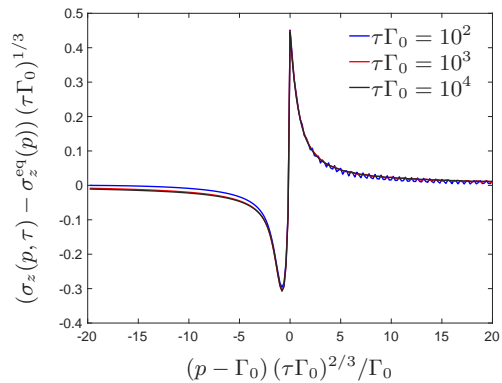}
\caption{Momentum resolved defect density for the non-hermitian drive.
The scaling of the numerically determined momentum resolved defect density, $f_\textmd{nh}(x)$ in the near adiabatic limit is shown
for several values of $\tau$ for the full non-hermitian drive around the equilibrium EP.
}
\label{scalingcase4}
\end{figure}

By taking a closer look at near-adiabatic dynamics, we calculate the momentum resolved spin expectation value, $\sigma_z(p,\tau)$ numerically. This is
 illustrated, together with its critical scaling, 
in Fig. \ref{scalingcase4}.
There is a clear difference in the
contribution of states with imaginary as opposed to real instantaneous eigenvalues to this expectation value.
This also differs significantly from the Landau-Zener transition probability of the corresponding hermitian system\cite{dziarmaga}. 
From the numerical data, the defect density obeys the scaling function
\begin{gather}
\sigma_z(p,\tau)=\sigma_z^\textmd{eq}(p)+\frac{1}{(\tau\Gamma_0)^{1/3}}f_\textmd{nh}\left(\frac{(p-\Gamma_0)^{1/2}}{\Gamma_0^{1/2}}(\tau\Gamma_0)^{1/3} \right)
\label{scalingfunc4}
\end{gather}
with $f_\textmd{nh}(x)$ the corresponding universal scaling function, depicted in Fig. \ref{scalingcase4} and $\sigma_z^\textmd{eq}(p)=$Re$\sqrt{1-(p/\Gamma_0)^2}$. 
Upon integrating this with respect to $p$, the $1/\tau$ scaling of the defect density follows.

This allows us to conjecture that for a general EP, the momentum resolved defect density satisfies the same form as Eq.
\eqref{scalingfunc3gen}, which agrees with Eq. \eqref{scalingfunc4} with $z=1/2$. Therefore, the induced defect density vanishes as $\tau^{-(d+z)\nu/(z\nu+1)}$ 
upon traversing the EP adiabatically, similarly
to the PT-symmetric ramp.

\section{Discussion}

Throughout this work, a non-hermitian Hamiltonian was used to generate the time evolution. From a more direct physical 
perspective, such dynamics follow from a Lindblad master equation in combination with a continuous 
measurement~\cite{carmichael,daley,ashida}. More specifically, consider a system, described just by the 
hermitian part of our Hamiltonian, coupled to an environment inducing radiative decay in  the 
individual two-level systems for each momentum $p$ with a rate $\Gamma$. In terms of a Lindblad equation, this results in quantum jump operators equal to $\sigma^-$ describing an incoherent decay upon emitting a photon. 
Equivalently, one can map the Lindblad dynamics for the system's density matrix onto a quantum jump trajectory picture, where pure states are evolved upon averaging over trajectories according to the following prescription. A single trajectory is specified by a set of quantum jumps at times, that can be sampled from a given probability distribution. At those times the quantum jump operator of the related Lindblad master equation is applied onto the quantum state. In our case this is the $\sigma^-$ operator and a quantum jump event can therefore be identified with the emission of a photon. In between the quantum jumps the dynamics is solely given by a non-hermitian Hamiltonian, which in our case is the one in Eq.~\eqref{hamilton}. 
In order to only select those trajectories without any photon emission event, which is the non-hermitian evolution targeted in this work, we can further continuously monitor the system and measure the number of emitted photons. In case we only consider those realizations, where no emission event has 
taken place, we end up with an evolution precisely captured by our non-hermitian Hamiltonian. Note that the measurement modifies the state of our system, since the absence of emission events effectively gradually forces the system over time towards 
the ground state of the two-level system as otherwise an emission event becomes too likely~\cite{daley}. This continuous 
measurement further ensures that the wave function is always properly normalized. While the 
anti-hermitian contribution forces the system towards the ground state of the two-level system, the 
hermitian part counteracts this tendency by coherently transferring population back into the excited state. At an EP, these processes compete most strongly and generate a nontrivial attractor of the dynamics: this is the single eigenstate of the 
non-hermitian system. This provides the additional channel for defect annihilation that we identified in our analysis. Far away from the EP, the 
hermitian part of the Hamiltonian dominates for our setups, so that the dynamics is almost fully coherent. 
Finally, we note that a distinct non-hermitian Kibble Zurek scaling describing a different physical realization with 
different definition of the expectation value and the scalar product was studied in Ref. \cite{shuaiyin}.

Single particle Hamiltonians of the form of Eq. \eqref{hamilton} have already been realized in non-conservative classical and quantum systems\cite{zeuner,ptreview,gao2015,xiao,gao2015}.
The time dependent control of the non-hermitian term together with measuring the spin components are possible. 
By creating several copies of this two level system, corresponding to distinct $p$'s, the effective many-body dynamics and the non-hermitian Kibble-Zurek scaling could in principle be detected.

To sum up, the universal features of non-hermitian dynamics across EPs were investigated. 
We find that the adiabatic time evolution drives the initial wavefunction to a right eigenstate of the final non-hermitian Hamiltonian, up to normalization, 
indicating that an adiabatic theorem probably exists for the systems under consideration. 
For a near adiabatic crossing of an EP, defects are produced at a reduced rate, whose density obeys a generalized Kibble-Zurek scaling as $\tau^{-(d+z)\nu/(z\nu+1)}$.
For the future it remains an open question how our results extend also to higher-order exceptional points.

\section{Methods}

\subsection{Non-hermitian time evolution}

For the purpose of this work we consider time-dependent parameters $\Delta(t)$ and/or $\Gamma(t)$ yielding a Hamiltonian $H(t)$.
Initially, before we start our parameter ramps, we choose the Hamiltonian always to be Hermitian, i.e. $\Gamma = 0$, so that the initial condition as the \emph{ground state} of the Hamiltonian is well-defined.
At time $t=0$ we start our time-dependent protocol over a time span $\tau$.
The time evolution follows from
\begin{gather}
i \partial_t | \Psi(t) \rangle = H(t) | \Psi(t) \rangle,
\label{timedepsch}
\end{gather}
with $|\Psi(t)\rangle = \otimes_p |\Psi_p(t) \rangle$ for a given mode $p$.
In general, the norm of the wave function is not conserved when time evolution is driven by a non-hermitian Hamiltonian, so that an additional prescription for performing measurements in such states has to be given.
When interpreting such dynamics as a result of dissipation in the framework of a Lindblad master equation with an additional continuous measurement, expectation values of an operator $\mathcal{O}$ have to be evaluated as\cite{ashida18,graefe2008,carmichael}
\begin{gather}
\langle \mathcal{O}(t)\rangle=\frac{\langle\Psi (t)|  \mathcal{O} |\Psi(t)\rangle}{\langle\Psi(t)|\Psi(t)\rangle} \, ,
\label{expvalue}
\end{gather}
where the left state, $\langle\Psi (t)|$ is taken as the hermitian conjugate of the time evolved right state, $|\Psi(t)\rangle$.
Since the initial condition at $t=0$ is chosen to be the ground state of a hermitian system, the initial right and left states also satisfy this condition.
In the following we will quantify the defect production via
\begin{equation}
        \langle \sigma_\alpha(t) \rangle = \frac{1}{N} \sum_p \sigma_\alpha(p), \,\,  \sigma_\alpha(p) =\frac{\langle\Psi_p (t)|  \sigma_\alpha  |\Psi_p(t)\rangle}{\langle\Psi_p(t)|\Psi_p(t)\rangle}
\label{expvalue1}
\end{equation}
with $\alpha=y,z$ and $N$ denoting the number of considered momentum states.
In the absence of balanced gain and loss, one can shift  the diagonal term in the Hamiltonian by any complex value, which does not affect the results.
The reason is that such a shift leaves the expectation values in Eq. \eqref{expvalue1} invariant, since a simple (time-dependent) change of the norm 
of the wave function is cancelled by explicitly using normalized expectation values.

\subsection{Calculation of the defect density}

The total defect density for the four considered settings are evaluated from the momentum resolved defect density after momentum integration. For example, by taking
the PT-symmetric ramp, Eq. \eqref{scalingfunc3} defines the momentum resolved defect density.
The total defect density is
\begin{gather}
\sigma_z(\tau)=\int \frac{dp}{2\pi} \sigma_z(p,\tau).
\end{gather}
By changing variable $x=p(\tau\Delta_0)^{1/3}/\Delta_0$, this becomes
\begin{gather}
\sigma_z(\tau)=\frac{1}{(\tau\Delta_0)^{1/3}}\frac{\Delta_0}{(\tau\Delta_0)^{1/3}}\int \frac{dx}{2\pi} f_\textmd{PT}(x)\sim \tau^{-2/3}.
\end{gather}

\begin{acknowledgments}
This research is supported by the National Research, Development and Innovation Office - NKFIH   within the Quantum Technology National Excellence Program (Project No.
      2017-1.2.1-NKP-2017-00001), K119442 and by
 UEFISCDI, project number  PN-III-P4-ID-PCE-2016-0032. In addition support by the Deutsche Forschungsgemeinschaft via the Gottfried  Wilhelm  Leibniz  Prize  program is gratefully acknowledged.
\end{acknowledgments}

\bibliographystyle{apsrev}
\bibliography{refgraph}

\end{document}